\documentclass{iopart}

\usepackage[hypertex]{hyperref}
\usepackage{srctex}
\usepackage{graphicx}
\usepackage{iopams}
\usepackage{amssymb}
\newcommand{\RS}{\scriptscriptstyle{\rm RS}}
\newcommand{\RSB}{\scriptscriptstyle{\rm 1RSB}}
\newcommand{\RSBt}{\scriptscriptstyle{\rm 2RSB}}
\begin{document}

\title{Replica analysis of the generalized p-spin interaction glass model.}

\author{T.I. Schelkacheva}
\address{Institute for High Pressure Physics, Russian Academy of Sciences, Troitsk 142190,
Moscow Region, Russia}
\author{N.M. Chtchelkatchev}
\address{Institute for High Pressure Physics, Russian Academy of Sciences, Troitsk 142190,
Moscow Region, Russia}
\address{Argonne National Laboratory, Argonne, IL 60439,USA}
\address{Department of Theoretical Physics, Moscow Institute of Physics and Technology, 141700 Moscow, Russia}
\begin{abstract}
We investigate stability of replica symmetry breaking solutions in generalized $p$-spin models. It is shown that the kind of the transition
to the one-step replica symmetry breaking state depends not only on the presence or absence of the
reflection symmetry of the generalized ``spin''-operators $\hat{U}$ but on the number of interacting operators and their
individual characteristics.
\end{abstract}

\maketitle

\section{Introduction\label{Sec:Intro}}
In the study of spin glasses the central role plays the Sherrington-Kirkpatrick (SK) model~\cite{sk}.  It appeared as an attempt to describe unordered freezing of spins in
dilute magnetic systems with disorder and frustration. This problem was soon solved at the mean-field level. It was demonstrated that the glassy phase in the SK-model is characterized by the the full replica symmetry breaking~\cite{P,A,Mezard}. Later the ``$p$''-spin  model was introduced as a natural generalization of the SK-model; $p$ spins interact in this model at each point of the lattice. It was shown in Ref.~\cite{Gardner} that in this case of $p$-spin interactions when $(p\geq 3)$ the full replica symmetry breaking (FRSB) does not appear  at the glass transition temperature $T_{c}$ and the 1-step replica symmetry breaking (1RSB) solution is stable in contrast to the SK model with 2-spin interaction. In addition it was discovered that the glass order parameter in the $p$-spin model is discontinuous at $T_c$ contrary to the SK-model.

The use of $p$-spin glass models as models for understanding structural glasses was pioneered in Ref.~\cite{WolynesT,WolynesK}. It was shown that these models qualitatively describe many aspects of the glass transitions in liquids, e.g. two critical temperatures. The number of metastable states in these models is similar to that obtained in numerical modeling in liquids. The structure of the dynamical equations for the correlation functions of supercooled liquids in mode-coupling theory and for $p$-spin model are similar~\cite{Geotze,Geotzee}.

Till now $p$-spin glass model remains to be in the focus of intense investigations (see, e.g., the recent papers Ref.~\cite{PP,Z,Cr,L,Crr,A.Montanari}) since it is a good starting point to understand the physics of real glasses. Some aspects  are still far from being completely understood. We consider a generalization of the $p$-spin model of Ising spins where arbitrary diagonal operators $\hat{U}$ stand instead of Ising spins~\cite{Sc,ESc}. The operators have different meanings depending on the problem under study. For example, Ising spin should be replaced with the molecule multipole moment if freezing of the orientational order is the target of the investigation~\cite{Sc,ESc,Walasek,Schelkacheva}.

An important development of the p-spin model is the study of the quadrupole system with $J = 2$ with multiparticle interaction, $p=3$~\cite{Schelkacheva}. For the full molecule momentum, $J = 2$, and $J_z = \{0,\pm1,\pm2\}$, the operator of the axial quadrupolar moment of a molecule has the form, $\hat{U} =\frac{1}{3}(3{J^{2}}_{z}-6)$. Using the operator $\hat U$ instead of the spin in the p-spin model it is possible to describe the orientational glass phase observed for high pressures in solid molecular ortho-$D_{2}$ and para-$H_{2}$. Moreover, it is possible to observe orientational transitions in such systems, which consist of initially spherically symmetric molecules with $J = 0$, because the probability of the transition, $(J = 0) \rightarrow (J = 2)$, increases rapidly with the pressure. The computational results show that the glass state and the long-range orientational order coexist, which agrees with experiments~\cite{G}. This model well describes high pressures since there the interactions of more than two particles play an important role.

We consider now a generalization of the $p$-spin model in the following way: the role of spins play the operators satisfying the condition of reflection symmetry, $\Tr {\hat{U}}^{(2k+1)} = 0$ for any integer $k$. Reflection symmetry of the operators $\hat{U}$ leads to vanishing of a number of terms in the free energy, so that the replica symmetric (RS) solution for the order parameters is zero at high temperature. As the result the behavior of 1RSB solution for the order parameters is like in the ordinary p-spin model of Ising spins with $(p\geq 3)$~\cite{Gardner}.

If operators $\hat{U}$ do not have the reflection symmetry, $ \Tr{\hat{U}}^{(2k+1)} \neq 0 $, then the glass freezing scenario is absolutely different from the Ising $p$-spin case.
The characteristic properties of system develop themselves already in the replica symmetry (RS) approximation. The nonlinear
integral equation for the RS-glass order parameter simply has no trivial solutions at any temperature because the integrand is nonsymmetric due to the cubic terms in the free-energy expansion~\cite{Sc,ESc}. There is a smooth increase in the order parameters (both glass and regular) as the temperature decreases. Thats why 1RSB order parameters appear continually at the bifurcation point  $T_{0}$. [In some sense 1RSB solution behaves here like in ordinary p-spin model with spins but in the external field ~\cite{Oliveira,Gillin}.]

We found $m_{1}$ analytically in the branching point and expressed the result through RS-order parameters. If $m_{1}\leq1$ in the branching point then 1RSB solution has physical meaning near $T_{0}$. The quadruple glass with $J=1$ and  $J_{z} =\{0,\pm1\}$ (three particle interaction) is the simplest example of the system without the reflection symmetry. In this case $\hat{U}= {3J^{2}}_{z}-2$ is the quadrupolar moment of the molecule  (see Fig.~1 in Ref.~\cite{ Sc}). Then 1RSB solution appears to be stable and it branches continuously at the bifurcation point  $T_{0} = T_{\rm \RSB}$ and smoothly on cooling.

If in the branching point $m_{1}>1$ then 1RSB solution does not have physical meaning in the vicinity of $T_{0}$. We suggest below an illustrative example of this conjecture.
At other realizations of $\hat{U}$ operators with $ \Tr{\hat{U}}^{(2k+1)} \neq 0 $ the transition from RS to 1RSB does not take place at the bifurcation point $T_0$ where 1RSB-solution appears (see Fig.~\ref{fig1}). Formally 1RSB parameter $m_{1} >1$ at $T_{0}$, however only $m_{1}\leq1$ have physical sense.
While $m_{1} >1$ then 1RSB solution formally exists but it is unstable. The temperature $T_{\rm \RSB}$ where stable 1RSB appears coincides with the point where $m_{1}=1$. At this point, $F_{\rm \RS} = F_{\rm \RSB}$. When the temperature is decreased $m_{1}$ becomes smaller than one and the 1RSB solution leads to a larger (preferable) free energy than the RS solution.

Below we investigate the crossover from continuous to jumpwise behavior of the glass order parameters in generalized p-spin models using the bifurcation theory and analyze the stability. We show that in general analytical progress can be made in the bifurcation region. We expand the boundaries of the ``ordinary'' $p$-spin model and consider 1RSB solutions also for the pairwise interactions.

\section{ Generalized p-spin model \label{Sec:glass}}

\subsection{Main equations}
The Hamiltonian of the $p$-spin model in general looks like:
\begin{equation}
H=-\sum_{{i_{1}}\leq{i_{2}}...\leq{i_{p}}}J_{i_{1}...i_{p}}
\hat{U}_{i_{1}}\hat{U}_{i_{2}}...\hat{U}_{i_{p}}, \label{one}
\end{equation}
where $\hat{U}$ now is arbitrary diagonal operator with $\Tr\hat{U}=0 $, $N$ is the number of sites on the lattice, $i=1,2,...N$, and $p$ is the finite integer giving the number of interacting particles. The coupling strengths are independent random variables with a Gaussian distribution
\begin{equation}
P(J_{i_{1}...i_{p}})=\frac{\sqrt{N^{p-1}}}{\sqrt{p!\pi}
J}\exp\left[-\frac{(J_{i_{1}...i_{p}})^{2}N^{p-1}}{ p!J^{2}}\right]. \label{two}
\end{equation}

Using replica approach we can write in general the free energy averaged over disorder:
\begin{eqnarray}
\nonumber\fl\langle F\rangle_J/NkT=\lim_{n \rightarrow 0}\frac{1}{n}\max\left \{- \frac{t^2}{4}\sum_{\alpha}
(w^{\alpha})^{p} + \sum_{\alpha}\mu^{\alpha}
w^{\alpha} -\right.
\\ \left.
\frac{t^2}{4}\sum_{\alpha\neq\beta} (q^{\alpha\beta})^{p}+ \sum_{\alpha\neq\beta}\lambda^{\alpha\beta} q^{\alpha\beta}-
\ln\Tr_{\{U^{\alpha}\}}\exp \hat{\theta}\right\}.\label{free0}
\end{eqnarray}
where $t={J}/kT$ and
\begin{equation}
\hat{\theta}=
\sum_{\alpha\neq\beta}\lambda^{\alpha\beta}\hat{U}^{\alpha}\hat{U}^{\beta}+
 \sum_{\alpha}\mu^{\alpha}(\hat{U}^{\alpha})^2. 
 \end{equation}

The extremum in Eq.~(\ref{free0}) should be taken over the physical order parameters and over the corresponding Lagrange multipliers, $\lambda^{\alpha\beta}$ and $\mu^{\alpha}$.
So the saddle point conditions give the glass order parameter $q^{\alpha\beta}$
\begin{equation}\label{four}
q^{\alpha\beta}= \frac{\Tr\left[\hat{U}^{\alpha}\hat{U}^{\beta}
\exp\left(\hat{\theta}\right)\right]} {\Tr\left[\exp\left(\hat{\theta}\right)\right]} ,
\end{equation}
the auxiliary order parameter $w^{\alpha}$, the regular order parameter $x^{\alpha}$:
\begin{equation}\label{five}
w^{\alpha}= \frac{\Tr\left[(\hat{U}^{\alpha})^2 \exp\left(\hat{\theta}\right)\right]}
{\Tr\left[\exp\left(\hat{\theta}\right)\right]},\qquad
x^{\alpha}= \frac{\Tr\left[\hat{U}^{\alpha}\exp\left(\hat{\theta}\right)\right]}
{\Tr\left[\exp\left(\hat{\theta}\right)\right]},
\end{equation}
and the parameters
\begin{equation}\label{five11}
\lambda^{\alpha\beta}=\frac{t^2}{4}p(q^{\alpha\beta})^{(p-1)},\qquad
\mu^{\alpha}=\frac{t^2}{4}p(w^{\alpha})^{(p-1)}.
\end{equation}

To proceed with replica symmetry breaking procedure it is more convenient to rewrite Eq.~(\ref{free0}) in the form:
\begin{eqnarray}
\nonumber\fl\langle F\rangle_J/NkT=\lim_{n \rightarrow 0}\frac{1}{n}\max\left \{(p-1) \frac{t^2}{4}\sum_{\alpha}
(w^{\alpha})^{p} + \right.
\\\left.
(p-1)\frac{t^2}{2}\sum_{\alpha>\beta} (q^{\alpha\beta})^{p}-
\ln\Tr_{\{U^{\alpha}\}}\exp \hat{\theta}\right\}.\label{free}
\end{eqnarray}
where
\begin{equation}
\hat{\theta}=p\frac{t^2}{2}
\sum_{\alpha>\beta}(q^{\alpha\beta})^{(p-1)}\hat{U}^{\alpha}\hat{U}^{\beta}+p\frac{t^2}{4}
 \sum_{\alpha}{(w^{\alpha})}^{(p-1)}(\hat{U}^{\alpha})^2. \label{six}
 \end{equation}


Using the standard procedure (see, e.g., Ref.~\cite{Mezard}), we perform the first stage of the replica symmetry breaking
(1RSB) ($n$ replicas are divided into  $n/m_{1}$ groups with  $m_{1}$ replicas in each) and obtain the expression for the free energy. Order parameters are denoted by
$q^{\alpha \beta }= r_{1}$ if $\alpha $ and
$\beta $ are from different groups and $q^{\alpha \beta }= r_{1}+v_{1}$ if
$\alpha $ and $\beta $ belong to the same group. So
\begin{eqnarray}
\nonumber\fl F_{\rm \RSB}=-NkT\left\{m_{1} t^2(p-1)\frac{r_{1}^p}{4}+(1-m_{1})(p-1)
t^2\frac{(r_{1}+v_{1})^p}{4}-t^2(p-1)\frac{{w_{1}}^p}{4}+ \right.
\\
\left. \frac{1}{m_{1}}\int dz^G\ln
\int ds^G
\left[\Tr\exp\left(\hat{\theta}_{\RSB}\right)\right]^{m_{1}}\right\}.
\label{frs}
\end{eqnarray}
Here
\begin{eqnarray}
\nonumber\fl \hat{\theta}_{\rm \RSB}=\left.zt\sqrt{\frac{p{r_{1}}^{(p-1)}}{2}}\,\hat{U}+st\sqrt{\frac{p[{(r_{1}+v_{1})}^{(p-1)}-{r_{1}}^{(p-1)}]}{2}}\,\hat{U}+
\right.
\\
\left.
t^2\frac{p[{w_{1}}^{(p-1)}-{(r_{1}+v_{1})}^{(p-1)}]}{4}\hat{U}^2,\right.
\end{eqnarray}
and
\begin{equation}
\int dz^G = \int_{-\infty}^{\infty} \frac{dz}{\sqrt{2\pi}}\exp\left(-\frac{z^2}{2}\right).
\end{equation}
The extremum conditions for $F_{\rm \RSB}$ yield equations for the glass order parameters
$r_{1}$ and $v_{1}$,
the additional order parameter $w_{1}$, the regular order parameter $x_{1}$ and the parameter $m_{1}$ :

\begin{eqnarray}\label{18qrs}
r_{1}=\int dz^G\left\{ \frac{\int ds^G{\left[\Tr\exp\hat{\theta}_{\rm \RSB}\right]}^{(m_{1}-1)}\left[\Tr\hat{U} \exp\hat{\theta}_{\rm \RSB}\right]}
{\int ds^G{\left[\Tr\exp\hat{\theta}_{\rm \RSB}\right]}^{m_{1}}}\right\}^{2},
\\\nonumber\fl
v_{1}=
\int dz^G \frac{\int ds^G{\left[\Tr\exp\hat{\theta}_{\rm \RSB}\right]}^{(m_{1}-2)}{\left[\Tr{\hat{U} }\exp\hat{\theta}_{\rm \RSB}\right]}^{2}}
{\int ds^G{\left[\Tr\exp\hat{\theta}_{\rm \RSB}\right]}^{m_{1}}}   -
\\
\int dz^G\left\{ \frac{\int ds^G{\left[\Tr\exp\hat{\theta}_{\rm \RSB}\right]}^{(m_{1}-1)}\left[\Tr\hat{U} \exp\hat{\theta}_{\rm \RSB}\right]}{\int ds^G{\left[\Tr\exp\hat{\theta}_{\rm \RSB}\right]}^{m_{1}}}\right\}^{2},\label{1vrs}
\\\label{1prs}
w_{1}=
\int dz^G \frac{\int ds^G{\left[\Tr\exp\hat{\theta}_{\rm \RSB}\right]}^{(m_{1}-1)}\left[\Tr{\hat{U} }^{2}\exp\hat{\theta}_{\rm \RSB}\right]}
{\int ds^G{\left[\Tr\exp\hat{\theta}_{\rm \RSB}\right]}^{m_{1}}},
\\
x_{1}=
\int dz^G \frac{\int ds^G{\left[\Tr\exp\hat{\theta}_{\rm \RSB}\right]}^{(m_{1}-1)}\left[\Tr{\hat{U} }\exp\hat{\theta}_{\rm \RSB}\right]}
{\int ds^G{\left[\Tr\exp\hat{\theta}_{\rm \RSB}\right]}^{m_{1}}},
\end{eqnarray}
and
\begin{eqnarray}\nonumber\fl
m_{1}\frac{t^{2}}{4}(p-1)\left[{(r_{1}+v)}^{p}-{(r_{1})}^{p}\right]=- \frac{1}{m_{1}}\int dz^G\ln\int ds^G \left[\Tr\exp\hat{\theta}_{\rm \RSB}\right]^{m_{1}}+
\\
\int dz^G \frac{\int ds^G{\left[\Tr\exp\hat{\theta}_{\rm \RSB}\right]}^{m_{1}}\ln\left[\Tr\exp\hat{\theta}_{\rm \RSB}\right]}{\int ds^G{\left[\Tr\exp\hat{\theta}_{\rm \RSB}\right]}^{m_{1}}}.\label{31mrs}
\end{eqnarray}
If operators $\hat{U}$ do not have the reflection symmetry then the nontrivial solution for the regular order parameter $x$ appears in spite of the fact that $x$ is absent in $\hat{\theta}$ (since $J_0=\langle J_{i_{1}...i_{p}}\rangle=0$ in (\ref{two}))~\cite{Sc,ESc}.

The heat capacity can be expressed through the glass order parameters:
\begin{equation}
\frac{C_{\rm v(\RSB)}}{kN}=\frac{d}{d(1/ t)} \left[ t \frac{m_{1}r_{1}^p+(1-m_{1})(r_{1}+v_{1})^p-{w_{1}}^p}{2}\right].
\end{equation}

The corresponding expressions for the RS approximation can be easily obtained from the preceding formulas (\ref{18qrs})-(\ref{1prs})
by setting $v_{1} = 0$. For glass order parameter $q_{\rm \RS}$ we have:
\begin{equation}\label{0qrs}
q_{\rm \RS}=\int dz^G\left\{ \frac{\Tr\left[\hat{U} \exp\left(\hat{\theta}_{\rm \RS}\right)\right]}
{\Tr\left[\exp\left(\hat{\theta}_{\rm \RS}\right)\right]}\right\}^{2}.
\end{equation}
Here
\begin{equation}\label{1qrs}
\hat{\theta}_{\rm \RS}=zt\sqrt{\frac{p\,{q_{\rm \RS}}^{(p-1)}}{2}}\,\hat{U}+t^2\frac{p[{w_{\rm \RS}}^{(p-1)}-
{q_{\rm \RS}}^{(p-1)}]}{4}\hat{U}^2.
\end{equation}

\subsection{Stability of the mean-field solutions}
The stability of the saddle point solution can be tested by the investigation of the gaussian fluctuation contribution to the free energy near this solution. The mean field (saddle point) solution is stable while all the eigen modes of the fluctuation propagator are positive. The most important mode is the so-called replicone mode \cite{A,ESc} since only its sign is usually very sensitive to the replica symmetry breaking degree and to the temperature. For example, the replica symmetric solution is stable unless the corresponding replicon mode energy ${\lambda_{\rm (\RS) repl}}>0$. The RS-solution can break at the temperature $T_{0}$ determined by the equation $\lambda_{\rm (\RS) repl}=0$, where
\begin{equation}\label{lambdaRS}
\fl\lambda_{\rm (\RS) repl}= 1 - t^{2} \frac{p(p-1){q_{\rm \RS}}^{(p-2)}}{2}
\int dz^G \left\{\frac{\Tr\left(\hat{U}^2
e^{\hat{\theta}_{\rm \RS}}\right)} {\Tr e^{\hat{\theta}_{\rm \RS}}}-
\left[\frac{\Tr\hat{U} e^{\hat{\theta}_{\rm RS}}}
{\Tr e^{\hat{\theta}_{\rm RS}}}\right]^2\right\}^2.
\end{equation}
The equation $\lambda_{\rm (\RS) repl}=0$ can be obtained as the branching condition for Eq.~(\ref{1vrs}), i.e., as the condition that a
small solution with 1RSB can appear.

We see that the equation for the glass order parameter,  Eq.(\ref{0qrs}), contains $\Tr[\hat{U} e^{\hat{\theta}_{\rm RS}}]/\Tr e^{\hat{\theta}_{\rm RS}}$. If operators $\hat{U}$ have zero trace for all odd powers: $\Tr{\hat{U}}^{(2k+1)}=0$ for all integer $k$, then  Eq.~(\ref{0qrs}) always has trivial solution $q_{\RS}=0$.

In fact, since Eq.~(\ref{0qrs}) can have other positive solutions, the stability condition is very useful to get the physical one. The stable solution for $T\neq 0$ is the trivial solution that bifurcates. Then equation $\lambda_{\rm (\RS) repl}(t)=0$ is solvable only for $p=2$ at finite $t=t_{0}$. Only in this case the nontrivial 1RSB solution smoothly branches  for $T_{0}\neq0$ . For $p>2$ and $\Tr{\hat{U}}^{(2k+1)}=0$, we obtain $T_{0}=0$. So 1RSB solution at $T_{\rm \RSB}\neq0$ appears discontinuously at $m_{1}=1$.

On the other hand, in the case $\Tr{\hat{U}}^{(2k+1)}\neq0$, the high-temperature expansion of the equation for the order parameter $q_{\RS}$ does not give trivial solution.
Using the condition, $q_{\RS}\neq 0$,  we can find that $T_{0}\neq0$. Then the solutions with the unbroken symmetry may appear continuously not only for $p=2$ but also at any integer $p$. For example, the stable continuous 1RSB solution exists for $p=3$ when $\hat{U} =3{J_{z}}^{2}-2$ is the axial quadrupole moment in the subspace $J=1$ with $J_{z}=0,\pm1$ (see Fig.~1 in Ref.~\cite{Sc}). Below we prove these statements.

In order to get in general 1RSB solution near the bifurcation  point  $T_0$ where it is close to the RS-solution we expand the free energy (\ref{free})-(\ref{six}) up to the third order, assuming that the deviations $\delta q^{\alpha \beta}$ from $q_{\rm \RS}$ and $\rho$ from $w_{\rm \RS}$ are small.

We use the notation $\Delta F$ for the difference of the free energy $F_{\rm (\RSB)}$ from its replica symmetry part  ${F^{\rm (\RSB)}_{0}}$. So,
\begin{eqnarray}\nonumber
\fl\frac{\Delta F}{NkT}=\frac{t^2}{4}\frac{p(p-1)}{2}{q^{(p-2)}_{\RS}} \left[1-t^{2}W\right]\left\{-\left[r-(m_{1}-1)v_{1}\right]^{2}-{v_{1}}^{2}m_{1}(1-m_{1})\right\}-
\\\nonumber
\fl\frac{t^{4}}{2}L\left[r-(m_{1}-1)v_{1}\right]^{2}-
\left.t^{6}\left\{C\left[r-(m_{1}-1)v_{1}\right]^{3}+\right.\right.
\\\nonumber\fl \left.
D\left[r-(m_{1}-1)v_{1}\right]{v_{1}}^{2}m_{1}(m_{1}-1)-B_{3}{v_{1}}^{3}{m_{1}}^{2}(m_{1}-1)+
\right.
\\
\left. B_{4}{v_{1}}^{3} m_{1}(m_{1}-1)(2m_{1}-1)\right\}+\Psi(\rho) +...\label{10frs}
\end{eqnarray}
where $t=t_{0}+\Delta t$, $r_{1}=q_{\rm \RS}+r$, $w_{\rm \RSB}=w_{\rm \RS}+\rho$ and the expressions for
$W$, $L$, $C$, $D$, $B_3$, $B_4$ and $\Psi $ are some combinations of operators averaged over the RS-solution (see Appendices A and B). For example, the coefficient $L$ enters $\Delta F$ like $ \lim_{n \rightarrow 0}\frac{1}{n} {\sum_{\alpha,\beta,\delta}}^{'}\delta q^{\alpha\beta}\delta q^{\alpha\delta}$.

First we consider the case $\Tr{\hat{U}}^{(2k+1)}\neq0$. Then $L|_{t=t_{0}}\neq {0}$.
Using extremum conditions for the free energy (\ref{10frs})
and the inequality $L|_{t=t_{0}}\neq {0}$,  we obtain the branching condition
\begin{equation}
r-(m_{1}-1)v_{1}=0+o(\Delta t)^{2}.
\label{4099prs}
\end{equation}
This condition states that there is no linear term for the glass order parameters. There is no other linear term  because
$\left[1-t^{2}W\right]|_{t=t_{0}}=\lambda_{\rm \RS repl}|_{t=t_{0}}=0$ at the branch point.
Since the coefficients $A_{k}\mid_{t_{0}}$ in the expression for $\Psi(\rho)$ (Appendix B) are in general nonzero  then from the extremum condition we get, $\rho\sim [r-(m_{1}-1)v_{1}]$ and $\Psi(\rho)=0+o(\Delta t)^{4}$.
Finally, we get:
 \begin{eqnarray} \label{30prs}
 2m_{1}(1-m_{1})\Gamma\Delta
t &=&3t_{0}^{6}m_{1}(1-m_{1})\left[-B_{4}+m_{1}(-B_{3}+2B_{4})\right]v_{1}\,,
\\\nonumber
 (2m_{1}-1)\Gamma\Delta t &=& t_{0}^{6}\left\{(2m_{1}-1)\left[-B_{4}+m_{1}(-B_{3}+2B_{4})\right]+\right.
\\&&\left.
m_{1}(m_{1}-1)(-B_{3}+2B_{4})\right\}v_{1}
\,,\label{40prs}
\end{eqnarray}
where $\Gamma$ is given explicitly in Appendix B.

Here $ B_{3}$,  $B_{4}$ and $\Gamma$  are taken at  $T=T_0$. Then we find from (\ref{30prs}) and (\ref{40prs}) (the cases $m_{1}=0$ and $m_{1}=1 $ should be investigated separately, see, e.g., Ref.~\cite{GribovaT}):
\begin{equation}
m_{1}={{B_{4}}/{B_{3}}}.\label{50prs}
 \end{equation}
at the branch point $T_0$ where 1RSB-solution appears and
\begin{equation}
 r_{1}=q_{\rm \RS}+(m_{1}-1)v_{1},\qquad
v_{1}\backsim
\Delta t;\label{77prs}
\end{equation}
in the neighborhood of  $T_0$.

The coefficient of proportionality  (\ref{77prs}) depends only on RS-solution at $T_{0}$:
\begin{equation}
\fl v_{1}=\frac{p(p-1)}{2}\frac{q_{\rm \RS}^{(p-2)}}{{6B_{4}(1-m_{1}){t_{0}}^{5}}}
\left\{1+\frac{t_{0}(p-2)}{2}\frac{\dot{q_{\rm \RS}}}{q_{\rm \RS}}+
\frac{t_{0}^{4}p^{2}(p-1)}{4}q_{\rm \RS}^{(p-2)}\Upsilon\right\} \Delta t,
\label{490prs}
\end{equation}
where
\begin{equation}
\fl \Upsilon=\left[\left(w_{\rm \RS}^{(p-1)}+\frac{t_{0}(p-1)}{2}w_{\rm \RS}^{(p-2)}\dot{w_{\rm \RS}}\right)K_{1}+
\left(q_{\rm \RS}^{(p-1)}+\frac{t_{0}(p-1)}{2}q_{\rm \RS}^{(p-2)}\dot{q_{\rm \RS}}\right)K_{2}\right],
\label{495prs}
\end{equation}
where $\dot{q_{\rm \RS}}=\frac{d}{dt}q_{\rm \RS}$ and $\dot{w_{\rm \RS}}=\frac{d}{dt}w_{\rm \RS}$.
Expressions $K_{1}$ and $K_{2}$ are quite long and they are written in Appendix B.

1RSB solution appears smoothly in most cases from RS solution. However  if the order parameter in the brunching point takes the unphysical values $m_{1}>1$ then at temperature where
$m_{1}=1$ and physical (stable) 1RSB solution appears it is drastically different from the RS solution and as the result the jump singularity appears in the heat capacity (e.g. Ref.~\cite{Schelkacheva}). The free energy does not have the discontinuity in this case (see equation (\ref{frs})).

\begin{figure}[t]
  \center
  \includegraphics[width=0.7\textwidth]{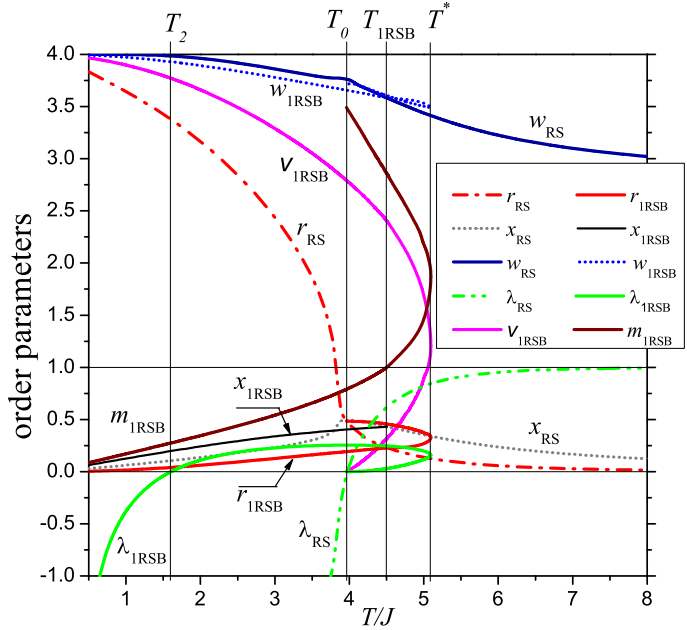}\\
  \caption{(Color online) Order parameters evolution with the temperature for 3-quadrupole model in the subspace $J = 2$. For simplicity only the physical solution for $x_{\rm \RSB}$ is shown here. There are four characteristic temperatures in the model: $T_2$, $T_0$, $T_{\rm \RSB}$ and $T^*$, where $\lambda_{\rm (\RSB) }=0$, $\lambda_{\rm (\RS) }=0$, $m_{\rm \RSB}=1$ and the branches of 1RSB order parameters merge correspondingly. There is region of temperatures in the graph, $T_0<T<T^*$, where two different 1RSB solutions coexist with the RS solution. The physical solution should have the largest free energy.  Our calculations show that 1RSB solution (with $m_{\rm \RSB}<1$)  has larger free energy than the RS solution when $T_0<T<T_{\rm \RSB}$. So this 1RSB solution is the physical one. The transition from the RS to the 1RSB solution occurs jumpwise at $T_{\rm \RSB}$. At this point, $F_{\rm \RS} = F_{\rm \RSB}$. So above $T_{\rm \RSB}$ the RS-solution is the physical one. When  $T_2<T<T_{\rm \RSB}$ then the 1RSB solution is also the physical one since it is stable with the respect to small perturbations having 2RSB symmetry.}
\label{fig1}
\end{figure}

As an example we consider the case of quadrupolar glass $\hat{U} = \hat{Q}=\frac{1}{3}\left[3{J_{z}}^{2}-6\right]$ \cite{Schelkacheva} in the subspace
$J=2$ in detail and write the explicit solutions of Eqs. (\ref{18qrs})–(\ref{31mrs}) for the number of interacting particles  $p=3$ (see Fig.~\ref{fig1}).
The glass state and the state with the long-range orientational order coexist. 1RSB solution appears smoothly from RS solution. But in this
case, the condition $\lambda_{\rm (\RS) repl}=0$ does not determine the physical solution in the neighborhood of the branch
point $T_{0}$, namely, a transition to the nonphysical branch of the free energy takes place. In fact, the transition from
the RS to the 1RSB solution occurs jumpwise at the point $T_{\rm \RSB} > T_{0}$ determined by the condition $m_{1} = 1$.
At this point, $F_{\rm \RS} = F_{\rm \RSB}$. Replica-symmetric solution is stable above $T_{\rm \RSB}$.  When the temperature is decreased $m_{1}$ becomes smaller than one and
the corresponding physical 1RSB solution corresponds to larger (preferable) free energy than the RS-solution. An exceptionally important property of the model is that there exists a domain of stability where the 1RSB-solution remains stable under the further RSBs (see below).

Now we shall consider the case $\Tr{\hat{U}}^{(2k+1)}=0$ and investigate 1RSB-solution near the branching point. Let us return to the expression (\ref{10frs}) for free energy at $p=2$ since only in this case the nontrivial 1RSB solution smoothly branch at finite $t$. Note that in the case of zero RS-solution for the glass order parameter the
expansion does not contain the terms where some indices occur only once. In the case of reflection symmetry operators there is no terms where some indices occur odd number of times. Отсюда  $L|_{t=t_{0}}= {0}$, (see. Appendix B), the branching condition (\ref{4099prs}) for 1RSB fails. Moreover at the branching point we have: $B_{4}=B'_{3}=B_{2}=B'_{2}=A_{3}= A_{12}=A_{14}={0}$; $C=2B_{3}$ и $D=-3B_{3}$. From the extremum condition we find $\Psi(\rho)=0+o(\Delta t)^{4}$ and
\begin{eqnarray}
m_{1}=0;
\\
r_{1}+v_{1}=\frac{1+{t_{0}}^{4}\left(w_{\RS}+\frac{t_{0}}{2}\dot{w}_{\RS}\right)\left[
\langle\hat{U}_{1}^{2}\hat{U}_{2}^{4}\rangle-
\langle\hat{U}_{1}^{2}\hat{U}_{2}^{2}\hat{U}_{3}^{2}\rangle\right]
}{\langle\hat{U}_{1}^{2}\hat{U}_{2}^{2}\hat{U}_{3}^{2}\rangle{t_{0}}^{5}}\Delta t.
\end{eqnarray}
It have been shown\cite{TSc} that the Parisi FRSB scheme can be used not only for the SK model but also for
any model with pair interactions and  with reflection symmetry Tr${\hat{U}}^{(2k+1)}=0$, for example, such as spin glasses with arbitrary spin.

We break the RS once more and obtain the corresponding expressions for the free energy and the order
parameters. The bifurcation condition  $\lambda_{\rm (\RSB) repl}=0$ determining the temperature $T = T_{2}$ follows from
the condition that a nontrivial small solution for the 2RSB glass order parameter appears as  $v_{2}\rightarrow 0$. We have:
\begin{eqnarray}
\nonumber\fl\lambda_{\rm (\RSB) repl}= 1 - t^{2} \frac{p(p-1)(r_{1}+v_{1})^{(p-2)}}{2} \times
\\
\fl\int dz^G \frac{\int ds^G\left[\Tr\exp\left(\hat{\theta}_{\RSB}\right)\right]^{m_{1}}
\left\{\frac{\Tr\left[\hat{U}^2
\exp\left(\hat{\theta}_{\RSB}\right)\right]} {\Tr\left[\exp\left(\hat{\theta}_{\RSB}\right)\right]}-
\left[\frac{\Tr\left[\hat{U} \exp\left(\hat{\theta}_{\RSB}\right)\right]}
{\Tr\left[\exp\left(\hat{\theta}_{\RSB}\right)\right]}\right]^2\right\}^2.}{\int ds^G
\left[\Tr\exp\left(\hat{\theta}_{\RSB}\right)\right]^{m_{1}}}\label{lambda}
\end{eqnarray}
Note that Eq.~(\ref{lambda}) depends only on  1RSB-solution. The expression for  $\lambda_{\rm (\RSB) repl}=0$ always has the solution for $v_{1}=0$, which determines the point $T_{0}$ and coincides with the solution of Eq.~(\ref{lambdaRS}) $\lambda_{\rm (\RS) repl}=0$ (see Fig.~\ref{fig1}).

Using the expressions for the glass order parameters obtained above near $T_{0}$ we can show that  $\frac{d}{dt}\lambda_{\rm (\RSB) repl}\mid_{T_{0}}=0$ as  for the two-body interaction in the presence of reflection symmetry and for the case$\Tr{\hat{U}}^{(2k+1)}\neq0$ and arbitrary $p$.

In addition to the point $T_{0}$, one more bifurcation point  $\lambda_{\rm (\RSB) repl}=0$ (see, e.g. Fig.~\ref{fig1}) may exist as  $v_{1}\neq0$,
and the 2RSB solution can appear at this point. At the point $T_{2}$ a transition to FRSB-state or to a stable
2RSB-state may take place.

We now present several results that hold in the two cases above. They concern the general form of the
expressions determining the stability of {1RSB} solutions.
Now we will investigate the stability of the 1RSB-solution to small perturbations having 2RSB symmetry for any value of the inverse temperature $t$. Group of replicas of $m_{1}$ elements, we divide by $m_{1}/m_{2}$ groups of $m_{2}$ elements each in order to find 2RSB solution.  The parameter $q^{\alpha\beta}$ in 2RSB case we shall denote $q_{2}^{\RSBt}$, if replicas $\alpha$ and $\beta$ belong the same and at the same time the smallest subgroup (amount of  $q_{2}^{\RSBt}$ elements is equal to $n(m_{2}-1)/2$). The elements of $q^{\alpha\beta}$ will be labeled by $q_{1}^{\RSBt}$ if replicase belong the same subgroup but this subgroup is not the smallest one (there are $n(m_{1}-m_{2})/2$ replicas of that kind). Finally the notation  $q_{0}^{\RSBt}$ will be used if the replicas $\alpha$ and $\beta$ belong to different subgroups [there are  $n(n-m_{1})/2$ replicas of this kind]. We set $q^{\alpha\beta} = {q^{\alpha\beta}}^{\RSB} + {\delta q^{\alpha\beta}}^{\RSBt}$ and assume that the deviations of the 1RSB solution from the 2RSB are small.
We believe that one can neglect the changes of the order parameter $w_{\RSB}$. Then the free energy  $\Delta F_{2}=F_{\RSBt}-F_{\RSB}$ can be conveniently represented as follows:
\begin{equation}\label{F2lambda}
\Delta F_{2}/NkT= \frac{t^{2}}{16}\left\{(\delta q_{0}^{\RSBt})^{2}a+(\psi_{2})^{2}b+(v_{2})^{2}c\right\},
\end{equation}
where the following substitution was used:
\begin{eqnarray}\label{F32lambda}
\delta q_{1}^{\RSBt}= \psi_{2}-\frac{d}{b}\delta q_{0}^{\RSBt}+(m_{2}-1)v_{2},
\\
\delta q_{2}^{\RSBt}= \psi_{2}-\frac{d}{b}\delta q_{0}^{\RSBt}+(m_{2}-m_{1})v_{2}.
\end{eqnarray}
Eq.(\ref{F2lambda})was found in Ref.~\cite{Gribova} for pairwise quadrupole interaction ($J=1$, $p=2$)without any assumptions about the behavior of the order parameter $w_{\RSB}$.

Parameters $a,b,c,d$ can be expressed through averages found on {1RSB}-solution. For example,
\begin{equation}\label{F432lambda}
\fl c=-2\lambda_{\rm (\RSB) repl}(m_{2}-m_{1})(1-m_{1})(1-m_{2})p(p-1)(r_{1}+v_{1})^{(p-2)}.
\end{equation}
Then we find that the parameter  $c\leq0$, if $\lambda_{\rm (\RSB) repl}\geq0$, because from Eqs.~(\ref{18qrs}),(\ref{1vrs}) follows that $(r_{1}+v_{1})>0$. Moreover the replica symmetry breaking scenario \cite{Mezard} assumes that  in the limit $n\rightarrow0$ we obtain $(m_{2}-m_{1})>0$,  $(1-m_{1})\geq0$ and $(1-m_{2})\geq0$.
Explicit form of the other parameters, namely, $a, b, d$, we give in the appendix (Appendix C). There it is shown that $b<0$ while $\lambda_{\rm (1RSВ) repl}>0$.

It is shown in Appendix C that while $\hat{U}$ satisfies the reflection symmetry condition then the coefficient $y_{1}\neq 0$ and the stability is determined by the sign of   $\lambda_{\rm (\RSB) repl}$ ($a<0$  if $\lambda_{\rm (\RSB) repl}>0$).

Direct numerical calculation in the subspace: $\hat{U}$ --- quadrupole operator with $J=2$, $p=3$, shows that $a<0$ if $\lambda_{\rm (\RSB) repl}>0$ and at the same time $(m_{1}-1)<0$, see Fig.~\ref{fig1}. Thus we can conclude that 1RSB solution is  stable indeed in this subgroup.

\section{Conclusions \label{Sec:Conc}}
To conclude, we have demonstrated that 1RSB solution in generalized $p$-spin models behaves differently depending on the symmetry of the operators $\hat{U}$. If the operators $\hat{U}$ have zero trace for all odd powers, $\Tr{\hat{U}}^{(2k+1)}=0$ for all integer $k$, then there trivial solution, $q_{\RS}=0$, for  RS order parameter always exists. Therefore the bifurcation condition, $\lambda_{\rm (\RS) repl}=0$, can be satisfied for finite $p\geq3$ only at $T_{0}=0$. In this case 1RSB solution smoothly branches only for $p=2$. When $p\geq3$ and $\Tr{\hat{U}}^{(2k+1)}=0$ 1RSB solution can appear only discontinuously at the temperature defined by the condition, $m_{1}=1$ (if $T_{\RSB}\neq0$) as in usual $p$-spin model. 1RSB solution is stable for $\lambda_{\rm (\RSB) repl}>0$.

On the contrary when $\Tr{\hat{U}}^{(2k+1)}\neq0$ solutions with broken replica symmetry may smoothly appear not only in the case of pair interactions, $p=2$, but also at any finite $p>2$. Point is that in this case the trivial RS-solution does not exist. The 1RSB branching condition $\lambda_{\rm (\RS) repl}=0$ is satisfied at finite temperature $T_{0}$. The solution with broken symmetry appears smoothly if  $m_{1}(T_0)\leq1$ otherwise physically stable 1RSB solution appears discontinuously at temperature where  $m_{1}=1$. Algebraical expression for $m_{1}(T_0)$ is obtained. It depends only on RS-solution at $T_{0}$.

\section{Acknowledgments}
Authors thank E.E. Tareyeva for active participation in the initial stage of the work and V.N. Ryzhov  for
helpful discussions and valuable comments.

This work was supported in part by the Russian Foundation for Basic Research (Grant No. 11-02-00341), by the President of the Russian Federation (Grant No. MK-7674.2010.2), the Russian Academy of Sciences programs and by the U.S. Department of Energy Office of Science through the contract DE-AC02-06CH11357.
\appendix
\section{}

The only nonzero sums:
\begin{eqnarray}\label{11eq:JP}
\fl\lim_{n \rightarrow 0}\frac{1}{n}
{\sum_{\alpha,\beta}}^{'}(\delta
q^{\alpha\beta})^{2}=-[r-(m_{1}-1)v_{1}]^{2}-m_{1}(1-m_{1}){v_{1}}^{2};
\\\label{51eq:JP}
\fl\lim_{n \rightarrow 0}\frac{1}{n}
{\sum_{\alpha,\beta,\delta}}^{'}\delta
q^{\alpha\beta}\delta
q^{\alpha\delta}=[r-(m_{1}-1)v_{1}]^{2};
\\\nonumber
\fl\lim_{n \rightarrow 0}\frac{1}{n}
{\sum_{\alpha,\beta}}^{'}(\delta
q^{\alpha\beta})^{3}=-[r-(m_{1}-1)v_{1}]^{3}+
3m_{1}(m_{1}-1)[r-(m_{1}-1)v_{1}]{v_{1}}^{2}+
\\
m_{1}(m_{1}-1)(2m_{1}-1){v_{1}}^{3};\label{61eq:JP}
\\\nonumber
\fl\lim_{n \rightarrow 0}\frac{1}{n}{\sum_{\alpha,\beta,\gamma}}^{'}\delta q^{\alpha\beta}\delta q^{\beta\gamma}\delta
q^{\gamma\alpha}=2[r-(m_{1}-1)v_{1}]^{3} - 3m_{1}(m_{1}-1)[r-(m_{1}-1)v_{1}]{v_{1}}^{2}-
\\
{m_{1}}^{2}(m_{1}-1){v_{1}}^{3};\label{711eq:JP}
\\\label{71eq:JP}
\fl \lim_{n \rightarrow
0}\frac{1}{n}{\sum_{\alpha,\beta,\gamma}}^{'}(\delta
q^{\alpha\beta})^{2}\delta
q^{\alpha\gamma}=[r-(m_{1}-1)v_{1}]^{3}-[r-(m_{1}-1)v_{1}]{v_{1}}^{2};
\\\label{81eq:JP}
\fl\lim_{n \rightarrow
0}\frac{1}{n}{\sum_{\alpha,\beta,\gamma,\delta}}^{'}\delta
q^{\alpha\beta}\delta q^{\alpha\gamma}\delta q^{\alpha\delta}=
\lim_{n \rightarrow
0}\frac{1}{n}{\sum_{\alpha,\beta,\gamma,\delta}}^{'}\delta
q^{\alpha\beta}\delta q^{\alpha\gamma}\delta q^{\beta\delta}=
-[r-(m_{1}-1)v_{1}]^{3}.
\end{eqnarray}

The prime on the sum means that only the superscripts belonging to the same
 $\delta q$ are necessarily different in  $\sum'$ .

\section{}

 \begin{eqnarray}\label{B11eq:JP}\fl W=\left[\frac{p(p-1)}{2}{q^{(p-2)}_{\RS}}\right]\left\{\langle\hat{U}_{1}^2\hat{U}_{2}^2\rangle-2\langle\hat{U}_{1}^2\hat{U}_{2}\hat{U}_{3}\rangle+
 \langle\hat{U}_{1}\hat{U}_{2}\hat{U}_{3}\hat{U}_{4}\rangle\right\};
 \\
 \fl L={\left[\frac{p(p-1)}{2}{q^{(p-2)}_{\RS}}\right]}^{2}\left\{\langle\hat{U}_{1}^2\hat{U}_{2}\hat{U}_{3}\rangle-
 \langle\hat{U}_{1}\hat{U}_{2}\hat{U}_{3}\hat{U}_{4}\rangle\right\};
 \\
 \fl C={\left[\frac{p(p-1)}{2}{q^{(p-2)}_{\RS}}\right]}^{3}\left\{-(B_{2}+B'_{2})+2B_{3}+B'_{3}-B_{4}\right\};
 \\
 \fl D={\left[\frac{p(p-1)}{2}{q^{(p-2)}_{\RS}}\right]}^{3}\left\{-3B_{3}-B'_{3}+3B_{4}\right\};
\end{eqnarray}
and
\begin{eqnarray*}\label{B12eq:JP}
\fl B_{2}={\left[\frac{p(p-1)}{2}{q^{(p-2)}_{\RS}}\right]}^{3}\left\{\frac{1}{2}\langle\hat{U}_{1}^{2}\hat{U}_{2}^{2}\hat{U}_{3}\hat{U}_{4}\rangle+ \frac{1}{2}
 \langle\hat{U}_{1}\hat{U}_{2}\hat{U}_{3}\hat{U}_{4}\hat{U}_{5}\hat{U}_{6}\rangle-
 \langle\hat{U}_{1}^{2}\hat{U}_{2}\hat{U}_{3}\hat{U}_{4}\hat{U}_{5}\rangle\right\};
\\
\fl B'_{2}={\left[\frac{p(p-1)}{2}{q^{(p-2)}_{\RS}}\right]}^{3}
\left\{\frac{1}{3}\langle\hat{U}_{1}\hat{U}_{2}\hat{U}_{3}\hat{U}_{4}\hat{U}_{5}\hat{U}_{6}\rangle-
 \frac{1}{2}\langle\hat{U}_{1}^{2}\hat{U}_{2}\hat{U}_{3}\hat{U}_{4}\hat{U}_{5}\rangle+\frac{1}{6}
 \langle\hat{U}_{1}^{3}\hat{U}_{2}\hat{U}_{3}\hat{U}_{4}\rangle\right\};
\\\nonumber
\fl B_{3}={\left[\frac{p(p-1)}{2}{q^{(p-2)}_{\RS}}\right]}^{3}\left\{\frac{1}{6}\langle\hat{U}_{1}^{2}\hat{U}_{2}^{2}\hat{U}_{3}^{2}\rangle-
\frac{1}{2}\langle\hat{U}_{1}^{2}\hat{U}_{2}^{2}\hat{U}_{3}\hat{U}_{4}\rangle-
\frac{1}{6}\langle\hat{U}_{1}\hat{U}_{2}\hat{U}_{3}\hat{U}_{4}\hat{U}_{5}\hat{U}_{6}\rangle+\right.
\\\left.
 \frac{1}{2}\langle\hat{U}_{1}^{2}\hat{U}_{2}\hat{U}_{3}\hat{U}_{4}\hat{U}_{5}\rangle\right\};
\end{eqnarray*}
\begin{eqnarray*}\nonumber
\fl t^{6}B'_{3}=t^{6}{\left[\frac{p(p-1)}{2}{q^{(p-2)}_{\RS}}\right]}^{3}\left\{-\langle\hat{U}_{1}\hat{U}_{2}\hat{U}_{3}\hat{U}_{4}\hat{U}_{5}\hat{U}_{6}\rangle+ \frac{5}{2}\langle\hat{U}_{1}^{2}\hat{U}_{2}\hat{U}_{3}\hat{U}_{4}\hat{U}_{5}\rangle-
 \frac{1}{2}\langle\hat{U}_{1}^{3}\hat{U}_{2}\hat{U}_{3}\hat{U}_{4}\rangle- \right.
\\
\fl \left.
\frac{3}{2}\langle\hat{U}_{1}^{2}\hat{U}_{2}^{2}\hat{U}_{3}\hat{U}_{4}\rangle+
 \frac{1}{2}\langle\hat{U}_{1}^{3}\hat{U}_{2}^{2}\hat{U}_{3}\rangle\right\}-
 \frac{t^{4}p^{2}{(p-1)}^{2}(p-2)}{8}{q^{(2p-5)}_{\RS}}\left\{-\langle\hat{U}_{1}^2\hat{U}_{2}\hat{U}_{3}\rangle+
 \langle\hat{U}_{1}\hat{U}_{2}\hat{U}_{3}\hat{U}_{4}\rangle\right\};
 \end{eqnarray*}
\begin{eqnarray*}\nonumber
\fl t^{6}B_{4}=t^{6}{\left[\frac{p(p-1)}{2}{q^{(p-2)}_{\RS}}\right]}^{3}\left\{\frac{1}{3}\langle\hat{U}_{1}\hat{U}_{2}\hat{U}_{3}\hat{U}_{4}\hat{U}_{5}\hat{U}_{6}\rangle- \langle\hat{U}_{1}^{2}\hat{U}_{2}\hat{U}_{3}\hat{U}_{4}\hat{U}_{5}\rangle+\frac{1}{3}\langle\hat{U}_{1}^{3}\hat{U}_{2}\hat{U}_{3}\hat{U}_{4}\rangle+ \right.
\\
\fl\left.
 \frac{3}{4}\langle\hat{U}_{1}^{2}\hat{U}_{2}^{2}\hat{U}_{3}\hat{U}_{4}\rangle-\frac{1}{2}\langle\hat{U}_{1}^{3}\hat{U}_{2}^{2}\hat{U}_{3}\rangle+
 \frac{1}{12}\langle\hat{U}_{1}^{3}\hat{U}_{2}^{3}\rangle\right\}-
 \frac{t^{2}p(p-1)(p-2)}{12}{q^{(p-3)}_{\RS}}\left[1-t^{2}\frac{3}{2}W\right];
 \end{eqnarray*}
 \begin{eqnarray*}
 \fl K_{1}=\langle\hat{U}_{1}^{2}\hat{U}_{2}^{4}\rangle-
 \langle\hat{U}_{1}^{2}\hat{U}_{2}^{2}\hat{U}_{3}^{2}\rangle-
 \langle\hat{U}_{1}^4\hat{U}_{2}\hat{U}_{3}\rangle-
 2\langle\hat{U}_{1}^{3}\hat{U}_{2}^{2}\hat{U}_{3}\rangle
 \\
 +3\langle\hat{U}_{1}^{2}\hat{U}_{2}^{2}\hat{U}_{3}\hat{U}_{4}\rangle+
 2\langle\hat{U}_{1}^{3}\hat{U}_{2}\hat{U}_{3}\hat{U}_{4}\rangle-
 2\langle\hat{U}_{1}^{2}\hat{U}_{2}\hat{U}_{3}\hat{U}_{4}\hat{U}_{5}\rangle;
 \end{eqnarray*}
 \begin{eqnarray*}
\fl K_{2}= \langle\hat{U}_{1}^{3}\hat{U}_{2}^{3}\rangle-8\langle\hat{U}_{1}^{3}\hat{U}_{2}^{2}\hat{U}_{3}\rangle+
21\langle\hat{U}_{1}^{2}\hat{U}_{2}^{2}\hat{U}_{3}\hat{U}_{4}\rangle+
6\langle\hat{U}_{1}^{3}\hat{U}_{2}\hat{U}_{3}\hat{U}_{4}\rangle-
\\
\fl 2\langle\hat{U}_{1}^{2}\hat{U}_{2}^{2}\hat{U}_{3}^{2}\rangle+
10\langle\hat{U}_{1}\hat{U}_{2}\hat{U}_{3}\hat{U}_{4}\hat{U}_{5}\hat{U}_{6}\rangle-
28\langle\hat{U}_{1}^{2}\hat{U}_{2}\hat{U}_{3}\hat{U}_{4}\hat{U}_{5}\rangle;
\end{eqnarray*}
\begin{eqnarray}\label{B1166eq:JP}
\fl A_{2}=\langle\hat{U}_{1}^2\hat{U}_{2}^{2}\rangle-
 \langle\hat{U}_{1}^{4}\rangle;
\\
\fl A_{3}= \langle\hat{U}_{1}^3\hat{U}_{2}\rangle-
 \langle\hat{U}_{1}^{2}\hat{U}_{2}\hat{U}_{3}\rangle;
\\
\fl A_{11} =\langle\hat{U}_{1}^6\rangle-
 3\langle\hat{U}_{1}^{4}\hat{U}_{2}^{2}\rangle+
2\langle\hat{U}_{1}^2\hat{U}_{2}^{2}\hat{U}_{3}^{2}\rangle;
\\
\fl A_{12}=\langle\hat{U}_{1}^{4}\hat{U}_{2}\hat{U}_{3}\rangle+
4\langle\hat{U}_{1}^{3}\hat{U}_{2}^{2}\hat{U}_{3}\rangle-
\langle\hat{U}_{1}^{3}\hat{U}_{2}^{3}\rangle-
3\langle\hat{U}_{1}^{2}\hat{U}_{2}^{2}\hat{U}_{3}\hat{U}_{4}\rangle-
\langle\hat{U}_{1}^{5}\hat{U}_{2}\rangle;
\\
\fl A_{13}=-\langle\hat{U}_{1}^{2}\hat{U}_{2}^{2}\hat{U}_{3}^{2}\rangle+
\langle\hat{U}_{1}^{4}\hat{U}_{2}^{2}\rangle+
3\langle\hat{U}_{1}^{2}\hat{U}_{2}^{2}\hat{U}_{3}\hat{U}_{4}\rangle-
\\
2\langle\hat{U}_{1}^{3}\hat{U}_{2}^{2}\hat{U}_{3}\rangle-
\langle\hat{U}_{1}^{4}\hat{U}_{2}\hat{U}_{3}\rangle
-2\langle\hat{U}_{1}^{2}\hat{U}_{2}\hat{U}_{3}\hat{U}_{4}\hat{U}_{5}\rangle+
2\langle\hat{U}_{1}^{3}\hat{U}_{2}\hat{U}_{3}\hat{U}_{4}\rangle;
\\\nonumber
\fl A_{14}=
3\langle\hat{U}_{1}^{2}\hat{U}_{2}^{2}\hat{U}_{3}\hat{U}_{4}\rangle-
2\langle\hat{U}_{1}^{3}\hat{U}_{2}^{2}\hat{U}_{3}\rangle-\\
\langle\hat{U}_{1}^{4}\hat{U}_{2}\hat{U}_{3}\rangle
-4\langle\hat{U}_{1}^{2}\hat{U}_{2}\hat{U}_{3}\hat{U}_{4}\hat{U}_{5}\rangle+
4\langle\hat{U}_{1}^{3}\hat{U}_{2}\hat{U}_{3}\hat{U}_{4}\rangle.
 \end{eqnarray}
\begin{eqnarray}\nonumber
\fl \Psi(\rho)=\rho^{2}\frac{t^2}{4}\frac{p(p-1)}{2}{w^{(p-2)}_{\RS}}\left[1+
\frac{p(p-1)}{2}{w^{(p-2)}_{\RS}}t^2A_{2}\right]+
\\\nonumber
\rho\,[r-(m_{1}-1)v_{1}]{\left[\frac{p(p-1)}{2}\right]}^{2}{w^{(p-2)}_{\RS}}{q^{(p-2)}_{\RS}}\frac{t^{4}}{2}A_{3}+
\\\nonumber
 \rho^{3}\frac{p(p-1)(p-2)}{3}{w^{(p-3)}_{\RS}}\frac{t^{2}}{4}\left[1+
\frac{3p(p-1)}{8}{q^{(p-2)}_{\RS}}t^2A_{2}\right]+
\\\nonumber
\rho^{2}[r-(m_{1}-1)v_{1}]\left[\frac{p(p-1)}{2}\right]^{2}(p-2)\frac{t^{4}}{2}
{w^{(p-2)}_{\RS}}{q^{(p-2)}_{\RS}}A_{3}+
\\\nonumber
\rho\,\left\{[r-(m_{1}-1)v_{1}]^{2}+v^{2}m_{1}(1-m_{1})\right\}\left[\frac{p(p-1)}{2}\right]^{2}
\frac{t^{4}}{2}
{w^{(p-2)}_{\RS}}{q^{(p-3)}_{\RS}}A_{3}-
\\\nonumber
\rho^{3}\frac{t^{6}}{48}\left[\frac{p(p-1)}{2}{w^{(p-2)}_{\RS}}\right]^{3}A_{11}-
\\\nonumber
\rho^{2}[r-(m_{1}-1)v_{1}]\frac{t^{6}}{8}\left[\frac{p(p-1)}{2}\right]^{3}
{w^{2(p-2)}_{\RS}}{q^{(p-2)}_{\RS}}A_{12}+
\\\nonumber
\rho\,\left\{[r-(m_{1}-1)v_{1}]^{2}+v^{2}m_{1}(1-m_{1})\right\}\frac{t^{6}}{4}
\left[\frac{p(p-1)}{2}\right]^{3}{w^{(p-2)}_{\RS}}{q^{2(p-2)}_{\RS}}A_{13}+
\\
\rho\, [r-(m_{1}-1)v_{1}]^{2}\frac{t^{6}}{4}
\left[\frac{p(p-1)}{2}\right]^{3}{w^{(p-2)}_{\RS}}{q^{2(p-2)}_{\RS}}A_{14};\label{B122eq:JP}
 \end{eqnarray}
\begin{equation}\label{B123eq:JP}
\Gamma = \frac{d}{dt}\left[\frac{t^2}{4}\frac{p(p-1)}{2}{q^{(p-2)}_{\RS}}\left[1-t^{2}W\right] \right],
\end{equation}
where
\begin{equation}
\langle\hat{U}_{1}^k\hat{U}_{2}^n...\rangle=\frac{\Tr\left[\hat{U}_{1}^k\hat{U}_{2}^n... \exp\Xi\right]}
{\Tr\left[\exp\Xi\right]} \label{0prs},
\end{equation}
and
\begin{equation}
\hat{\Xi}=\frac{t^{2}}{4}p{w_{\RS}}^{(p-1)}\sum_{\alpha}(\hat{U}^{\alpha})^{2}+
\frac{t^{2}}{4}p{q_{\RS}}^{(p-1)}\sum_{\alpha\neq\beta}\hat{U}^{\alpha}\hat{U}^{\beta}.
\end{equation}

\begin{center}

\section{}
\end{center}
\begin{eqnarray}\nonumber
\fl b=-(1-m_{1}) 2\lambda_{\rm (1 RSB) repl}p(p-1)(r_{1}+v_{1})^{(p-2)}-
\\\nonumber
\fl (1-m_{1})^{2} \frac{t^{2}}{2}p^{2}(p-1)^{2}(r_{1}+v_{1})^{2(p-2)}
\left\{4\int dz^G \frac{\int ds^G\left[{\Tr e^{\hat{\theta}_{1RSB}}}\right]^{m_{1}}
\frac{\Tr\left(\hat{U}^2
e^{\hat{\theta}_{1RSB}}\right)} {\Tr e^{\hat{\theta}_{1RSB}}}
\left[\frac{\Tr\left(\hat{U}
e^{\hat{\theta}_{1RSB}}\right)} {\Tr e^{\hat{\theta}_{1RSB}}}\right]^2}{\int ds^G
\left[{\Tr e^{\hat{\theta}_{1RSB}}}\right]^{m_{1}}}+\right.
\\\nonumber
\fl \left.
(m_{1}-4)\int dz^G \frac{\int ds^G\left[{\Tr e^{\hat{\theta}_{1RSB}}}\right]^{m_{1}}
\left[\frac{\Tr\left(\hat{U}
e^{\hat{\theta}_{1RSB}}\right)} {\Tr e^{\hat{\theta}_{1RSB}}}\right]^4}{\int ds^G
\left[{\Tr e^{\hat{\theta}_{1RSB}}}\right]^{m_{1}}}-\right.
\\\left.
m_{1}\int dz^G \left[\frac{\int ds^G\left[{\Tr e^{\hat{\theta}_{1RSB}}}\right]^{m_{1}}
\left[\frac{\Tr\left(\hat{U}
e^{\hat{\theta}_{1RSB}}\right)} {\Tr e^{\hat{\theta}_{1RSB}}}\right]^{2}}{\int ds^G
\left[{\Tr e^{\hat{\theta}_{1RSB}}}\right]^{m_{1}}}\right]^{2}\right\}.
\end{eqnarray}

We shall further use the following notations:
$\hat{\theta}_{\RSB}=\Theta$ and  $\langle J_{0}\rangle =\int ds^G{\left[\Tr {e}^{\Theta}\right]}^{m_{1}}$.

It follows from the Cauchy--Schwarz inequality that the expression $\left(\Tr\hat{U}^{2}e^{\Theta}\right)\left(\Tr e^{\Theta}\right)\geq\left(\Tr \hat{U}e^{\Theta}\right)^{2}$ follows from $\left(\sum_{n}{A_{n}}^{2}\right)\left(\sum_{n}{B_{n}}^{2}\right)\geq\left(\sum_{n}A_{n}B_{n}\right)^{2} $. So
\begin{equation*}\fl
4\int\frac{dz^G}{\langle J_{0}\rangle}\int ds^G\left(\Tr e^{\Theta}\right)^{m_{1}-4}\left(\Tr \hat{U}e^{\Theta}\right)^{2}
\left\{\left(\Tr {\hat{U}}^{2}e^{\Theta}\right)\left(\Tr e^{\Theta}\right)-\left(\Tr {\hat{U}}e^{\Theta}\right)^{2}\right\}\geq 0.
\end{equation*}
Then we find similarly that $\int dx{A(x)}^{2}\int dx{B(x)}^{2}\geq[\int dx{A(x)}{B(x)}]^{2}$, and
\begin{eqnarray}\nonumber
\fl m_{1}\int dz^G\frac{1}{{\langle J_{0}\rangle}^{2}}\left\{\int ds^G\left(\Tr e^{\Theta}\right)^{m_{1}}\int ds_{1}^G\left(\Tr e^{\Theta}\right)^{m_{1}-4}\left(\Tr \hat{U}e^{\Theta}\right)^{4}-\right.
\\
\left.
\left[\int ds^G\left(\Tr e^{\Theta}\right)^{\frac{m_{1}}{2}}\left(\Tr e^{\Theta}\right)^{\frac{m_{1}}{2}-2}\left(\Tr \hat{U}e^{\Theta}\right)^{2}\right]^{2}\right\}\geq0.
\end{eqnarray}
Then it follows that $b<0$ when $\lambda_{\rm (\RSB) repl}>0$.

Here $a=(\tilde{a}b-d^{2})/b$, where
\begin{eqnarray}\fl\nonumber
\tilde{a}=-2m_{1}p(p-1){r_{1}}^{(p-2)}\left\{1-\frac{t^{2}}{2}p(p-1){r_{1}}^{(p-2)}\left[\left(y_{1}+y_{2} -2y_{3}\right)+\right.\right.
\\
\left.\left.
4m_{1}y_{4}-2m_{1}y_{3}-3{m_{1}}^{2}y_{5}-4m_{1}(1-m_{1})y_{6}+m_{1}(2-m_{1})y_{2}\right]\right\},
\end{eqnarray}
\begin{equation}\fl
d=-m_{1}(1-m_{1})t^{2}p^{2}(p-1)^{2}{r_{1}}^{(p-2)}{(r_{1}+v_{1})}^{(p-2)}\left\{2y_{7}-m_{1}y_{6}-(2-m_{1})y_{8}\right\},
\end{equation}
and
\begin{eqnarray*}
 y_{1}=\int dz^G\frac{1}{{\langle J_{0}\rangle}^{2}}\left[\int ds^G\left(\Tr e^{\Theta}\right)^{m_{1}-1}\left(\Tr {\hat{U}}^{2}e^{\Theta}\right)\right]^{2};
\\
y_{2}=\int dz^G\frac{1}{{\langle J_{0}\rangle}^{2}}\left[\int ds^G\left(\Tr e^{\Theta}\right)^{m_{1}-2}\left(\Tr \hat{U}e^{\Theta}\right)^{2}\right]^{2};
\\
\fl y_{3}=\int dz^G\frac{1}{{\langle J_{0}\rangle}^{2}}\int ds^G\left(\Tr e^{\Theta}\right)^{m_{1}-2}\left(\Tr \hat{U}e^{\Theta}\right)^{2}\int ds_{1}^G\left(\Tr e^{\Theta}\right)^{m_{1}-1}\left(\Tr {\hat{U}}^{2}e^{\Theta}\right);
\\
\fl y_{4}=\int dz^G\frac{1}{{\langle J_{0}\rangle}^{3}}\int ds_{1}^G\left(\Tr e^{\Theta}\right)^{m_{1}-1}\left(\Tr {\hat{U}}^{2}e^{\Theta}\right)
\left[\int ds^G\left(\Tr e^{\Theta}\right)^{m_{1}-1}\left(\Tr \hat{U}e^{\Theta}\right)\right]^{2};
\\
\fl y_{5}=\int dz^G\frac{1}{{\langle J_{0}\rangle}^{4}}\left[\int ds^G\left(\Tr e^{\Theta}\right)^{m_{1}-1}\left(\Tr \hat{U}e^{\Theta}\right)\right]^{4};
\\
\fl y_{6}=\int dz^G\frac{1}{{\langle J_{0}\rangle}^{3}}\int ds_{1}^G\left(\Tr e^{\Theta}\right)^{m_{1}-2}\left(\Tr {\hat{U}}e^{\Theta}\right)^{2}
\left[\int ds^G\left(\Tr e^{\Theta}\right)^{m_{1}-1}\left(\Tr \hat{U}e^{\Theta}\right)\right]^{2};
\\
\fl y_{7}=\int dz^G\frac{1}{{\langle J_{0}\rangle}^{2}}\int ds^G\left(\Tr e^{\Theta}\right)^{m_{1}-1}\left(\Tr \hat{U}e^{\Theta}\right)\int ds_{1}^G\left(\Tr e^{\Theta}\right)^{m_{1}-2}\left(\Tr {\hat{U}}e^{\Theta}\right)\left(\Tr {\hat{U}}^{2}e^{\Theta}\right);
\\
\fl y_{8}=\int dz^G\frac{1}{{\langle J_{0}\rangle}^{2}}\int ds^G\left(\Tr e^{\Theta}\right)^{m_{1}-3}\left(\Tr \hat{U}e^{\Theta}\right)^{3}\int ds_{1}^G\left(\Tr e^{\Theta}\right)^{m_{1}-1}\left(\Tr {\hat{U}}e^{\Theta}\right);
\end{eqnarray*}


\begin{thebibliography}{99}

\bibitem{sk} Sherrington D and Kirkpatrick S 1975 Phys. Rev. Lett. {\bf
32} 1972; Kirkpatrick S and Sherrington D 1978 Phys. Rev. B {\bf 17},
4384

\bibitem{P} Parisi G 1980 J. Phys. A {\bf 13} L115

\bibitem{A}Almeida J R L and Thouless D J 1978 J. Phys. A {\bf 11}, 983

\bibitem{Mezard} Mezard M., Parisi G. and Virasoro M. 1987 \textit{Spin Glass Theory and
beyond} (World Scientific, Singapore)

\bibitem{Gardner} Gardner E 1985  Nuc. Phys. {\bf B257}, 747

\bibitem{WolynesT} Kirkpatrick T R and Thirumalai D 1987 Phys.
Rev. B {\bf 36}, 5388

\bibitem{WolynesK} Kirkpatrick T R and
Wolynes P G 1987 Phys. Rev. B {\bf 36}, 8552

\bibitem{Geotze} Geotze W. 1991 \textit{Liquid, Freezing and the Glass Transition}
edited by Hansen J P , Levesque D and Zinn-
Justin J (Elsevier, New York)

\bibitem{Geotzee} Bouchaud J-P,
Cugliandolo L F, Kurchan J and Mezard M 1997 \textit{Spin Glasses
and Random Fields}, edited by Young A P (World Scientific:
Singapore)

\bibitem{PP} Parisi G 2003 \textit{Les Houches Summer School - Session LXXVII: Slow relaxation and non equilibrium dynamics in condensed matter} ed. by Barrat J L, Feigelman M V, Kurchan J and Dalibard J (Elsevier)

\bibitem{Z} Zamponi F http://arxiv.org/abs/1008.4844.


\bibitem{Cr} Crisanti A, Leuzzi L,Rizzo T 2005 Phys.Rev. B {\bf 71}, 094202

\bibitem{L} Crisanti A and Leuzzi L 2006 Phys. Rev. B {\bf 73}, 014412

\bibitem{Crr} Crisanti A and Leuzzi L 2007 Phys.Rev. B {\bf 76}, 184417


\bibitem{A.Montanari}Montanari A andRicci-Tersenghi F 2003 Eur. Phys. J. B {\bf 33}, 339


\bibitem{Sc} Schelkacheva T I, Tareyeva E E andChtchelkatchev N M 2010 Phys. Rev. B {\bf{82}}, 134208

\bibitem{ESc} Tareyeva E E, Schelkacheva T I and Chtchelkatchev N M 2009 Theor. Math. Phys. {\bf{160}}, 1190

\bibitem{Walasek} Walasek K 1995 J. Phys. A: Math. Gen. {\bf 28}, L497

\bibitem{Schelkacheva} Schelkacheva T I, Tareyeva E E and Chtchelkatchev N M 2009 Phys. Rev. E {\bf{79}}, 021105

\bibitem{G} Goncharov A F, Eggert J H, Mazin I I, Hemley R J and Mao H K 1996 Phys. Rev. B {\bf{54}}, R15590

\bibitem{Oliveira} Viviane M de Oliveira and Fontanari J F 1999 J. Phys. A: Math. Gen. {\bf{32}}, 2285

\bibitem{Gillin}  Gillin P. and Sherrington D 2001 J. Phys. A \textbf{34} 1219

\bibitem{tren} Vainberg M M and Trenogin V A 1974 {\it Theory of branching of solutions of
non-linear equations (Monographs and Textbooks on Pure and Applied
Mathematics)} (Leyden: Noordhoff International Publishing)

\bibitem{DD} Crisanti A and De Dominicis C 2010 J. Phys. A: Math. Theor. {\bf {43}}, 055002

\bibitem{GribovaT} Gribova N V, Schelkacheva T I and Tareyeva E E 2010 J. Phys. A: Math. Theor. {\bf {43}}, 495006

\bibitem{TSc} Schelkacheva T I, Tareyeva E E and Chtchelkatchev N M 2006 Phys. Lett. A {\bf{358}}, 222

\bibitem{Gribova} Gribova N V and Tareyeva E E 2002 Teor. Math. Phys. {\bf{131}}, 852

\end{thebibliography}
\end{document}